\documentclass[paper]{JHEP3}
\usepackage{epsfig}
\newcommand{\gev}{\mbox{GeV}}

  \newcommand{\ccaption}[2]{
    \begin{center}
    \parbox{0.85\textwidth}{
      \caption[#1]{\small{{#2}}}
      }
    \end{center}
 \vskip 0.3truecm
    }

\def    \be             {\begin{equation}}
\def    \ee             {\end{equation}}
\def    \beq            {\begin{equation}}
\def    \eeq            {\end{equation}}
\def    \ba             {\begin{eqnarray}}
\def    \ea             {\end{eqnarray}}
\def    \beqn           {\begin{eqnarray}}
\def    \eeqn           {\end{eqnarray}}
\def    \beeq           {\begin{eqnarray}}
\def    \eeeq           {\end{eqnarray}}

\def    \=              {\;=\;}
\def    \frac           #1#2{{#1 \over #2}}

\def\abs#1{\left|#1\right|}
\def \lsim{\mathrel{\vcenter
     {\hbox{$<$}\nointerlineskip\hbox{$\sim$}}}}
\def \gsim{\mathrel{\vcenter
     {\hbox{$>$}\nointerlineskip\hbox{$\sim$}}}}

\def    \bra#1          {\mbox{$\langle #1 |$}}
\def    \ket#1          {\mbox{$| #1 \rangle$}}

\def    \gev            {\mbox{$\mathrm{GeV}$}}

\def    \mw             {\mbox{$m_W$}}  
\def    \mwt            {\mbox{$m_W^2$}}  
\def    \gw             {\mbox{$\Gamma_W$}}  
\def    \mz             {\mbox{$m_Z$}}

\def    \pt             {\mbox{$p_T$}}
\def    \ptel           {\mbox{$p_T^e$}}
\def    \ptw            {\mbox{$p_T^W$}}
\def    \yw             {\mbox{$y_W$}}

\def    \etael             {\mbox{$\eta^e$}}
\def    \met            {\mbox{$\rlap{\kern.2em/}E_T$}}

\newcommand     \MSB            {\ifmmode {\overline{\rm MS}} \else 
                                 $\overline{\rm MS}$  \fi}
\def    \muf            {\mbox{$\mu_{\rm F}$}}

\def    \mur            {{\mbox{$\mu_{\rm R}$}}}

\def    \as             {\ifmmode \alpha_s \else $\alpha_s$ \fi}
\def    \asmz             {\ifmmode \alpha_s(M_Z) \else $\alpha_s(M_Z)$ \fi}

\def \oas   {\mbox{$ {\cal O}(\alpha_s)$}}

\def\rt1{\raisebox{-1ex}{\rlap{$\; \rho \to 1 \;\;$}}
\raisebox{.4ex}{$\;\; \;\;\simeq \;\;\;\;$}}

\def\herwig{{\small HERWIG}}

\def\ALPGEN{{\small ALPGEN}}
\def\ppbar{$p \bar{p}$}
\title{How accurately can we measure the $W$ cross section?}
\vfill                                                       
\author{S. Frixione \\ 
INFN, Sezione di Genova, Italy\\
  E-mail: \email{Stefano.Frixione@cern.ch}}
\author{M.L. Mangano \\
CERN, Theoretical Physics Division, Geneva, Switzerland\\
  E-mail: \email{Michelangelo.Mangano@cern.ch}}
\abstract{
We study the QCD sources of systematic uncertainties in the experimental
extraction of the $W$ cross section at hadron colliders. The uncertainties
appear in the evaluation of the detector acceptances used to convert the
number of observed events into a total production cross section. We
consider the effect of NLO corrections, as well as of the inclusion of
parton showers, and evaluate the impact of spin correlations and of PDF
and scale uncertainties.
}
\preprint{CERN--PH--TH/2004--081\\
          GEF--TH--6/2004\\
          hep-ph/0405130}

\begin{document}


\section{Introduction}
$W$ production, through its leptonic decays,  features one of the cleanest
signatures at hadronic colliders, with a high-$\pt$ charged lepton
recoiling against missing energy~\cite{Arnison:rp}--\cite{Abachi:1995xc}.
This distinctive signature and the large production rates allow 
the measurements of the $W$ mass ($\mw$) performed at the
Tevatron~\cite{Affolder:2000bp}--\cite{unknown:2003sv} to be competitive 
with LEP2 results; a further improvement is expected at the LHC. Accurate
measurements of the total $W$ width ($\gw$) will also be obtained.
The experimental techniques necessary to perform these measurements are 
well known~\cite{Albajar:1990hg,Alitti:1991dm}, 
and tested extensively at the Tevatron Run I: $\gw$ has been
extracted with ``indirect'' 
methods~\cite{Abe:1995wf,Abbott:1999tt}
(in which the measured quantity is the ratio of the
$Z$ over the $W$ cross section), and with ``direct'' 
methods~\cite{Affolder:2000mt,Abazov:2002xj} (in which the measured
quantity is the distribution of the $W$ transverse mass). In both cases, 
a firm control is mandatory on the theoretical predictions for the
$p\bar{p}\to W+X$ or $pp\to W+X$ production processes; the cross
sections for these can be schematically written as follows:
\beq
\sigma^{\rm th}(W)=\sum_{ab} {\cal P}_{ab}\otimes \hat{\sigma}_{ab}(W)\,.
\label{factth}
\eeq
Here, ${\cal P}_{ab}$ is the product of the
parton density functions (PDFs) of the partons $a$ and $b$ (quarks 
and gluons) in the colliding protons/antiprotons. The PDFs cannot
be computed in QCD at present, and are extracted from global fit to
data (dominated by DIS); on the other hand the quantity 
$\hat{\sigma}_{ab}(W)$, the cross section of the process $ab\to W+X$,
{\em is} theoretically computable. In fact, the overwhelming majority
of the theoretical work on $W$ production has the scope of improving 
the accuracy with which $\hat{\sigma}_{ab}(W)$ is known. NLO QCD corrections
have been computed a long time ago~\cite{Altarelli:1979ub}--\cite{Humpert:ux},
in a series of papers which pioneered the factorization techniques in
perturbation theory. Total rates to NNLO accuracy have been presented in
ref.~\cite{Hamberg:1990np}--\cite{Harlander:2002wh}; 
recently, the NNLO result for the rapidity
of the $W$ has also become available~\cite{Anastasiou:2003ds}, the first
differential distribution ever to be computed at this order in $\as$.
The resummation of the leading and next-to-leading logarithms of
$\ptw/\mw$, relevant to the small-$\ptw$ region where the previously
mentioned fixed-order results are unreliable, has been incorporated
in a code available to experiments~\cite{Balazs:1997xd}.
EW corrections to the $W$ cross section have been shown to be non
negligible~\cite{Baur:1998kt}--\cite{Dittmaier:2001ay}, with
effects  up to 5\%; fortunately, the dominant contribution there is 
due to photon emission, which can be implemented in Monte Carlos
with multiple QED radiation~\cite{CarloniCalame:2004qw}, or combined
with resummed QCD formulae~\cite{Cao:2004yy} (which is especially 
relevant to $\mw$ measurements).

The measurements of the $W$ mass or width are performed by solving for
$\mw$ or $\gw$ the following equation
\beq
\sigma^{\rm th}(W)=\sigma^{\rm exp}(W)\,,
\label{theqexp}
\eeq
where $\sigma^{\rm exp}(W)$ is the experimental result for the relevant
observable. This can be written as follows:
\beq
\sigma^{\rm exp}(W)=\frac{1}{{\rm BR}(W\to l\nu)}\,\frac{1}{\int\!{\cal L}dt}\,
\frac{N^{\rm obs}}{A_W}\,,
\label{sigexp}
\eeq
where ${\rm BR}(W\to l\nu)$ is the branching ratio for the leptonic decay
of the $W$ considered, $\int\!{\cal L}dt$ is the integrated hadron 
luminosity, $N^{\rm obs}$ is the number of detected signal events, 
and $A_W$ is the acceptance, namely the fraction of events which pass 
the selection cuts of the experimental analysis\footnote{We neglect 
for simplicity the discussion of the experimental {\it efficiency} with 
which events within the acceptance can be detected.}.
A procedure alternative to that of measuring the $W$ mass or 
width is that of using the world averages for $\mw$ and $\gw$,
and to solve eq.~(\ref{theqexp}) for $\int\!{\cal L}dt$; in this
way, $W$ production is treated as a (hadron) {\em luminosity monitor}.
The use of hard processes as luminosity monitors is a very interesting
possibility in the high-energy regime of the LHC, as opposed to the
more traditional determination of the luminosity through the 
knowledge of the total hadronic cross section; a necessary
condition for this to happen is that the hard cross section must be
reliably computed with small uncertainty. The procedure can be pushed
a step further, and eq.~(\ref{theqexp}) can be solved for 
${\cal P}_{ab}\int\!{\cal L}dt$, or for ${\cal P}_{ab}$; thus, $W$ production 
is used in the former case as a {\em parton} luminosity monitor, and in the
latter case to determine the PDFs. In both cases, the difficulty lies in
the sum over the parton labels appearing in eq.~(\ref{factth}): one 
needs to devise a way to force the particular combination of partons
he is interested in to be the dominant one in that sum. This always
implies the necessity of considering differential $W$ distributions.
A well known example is the determination of the ratio 
of the $d$ and $u$ parton densities, which is accessible through the 
rapidity distribution of the $W$. More details on the use of $W$ production  
as a luminosity monitor or in the context of PDF determination
can be found in refs.~\cite{Dittmar:1997md,Khoze:2000db,Giele:2001ms}.

Equation~(\ref{theqexp}) only involves $W$ cross sections. However, the $W$'s
are not detected as such by experiments, but only through charged leptons and
missing energy; furthermore, this detection necessarily takes place only in a
part of the final-state phase space, either because of limited detector
acceptance, or to avoid regions where the backgrounds are so large that the
determination of the signal is totally unreliable. Thus, the quantity 
$N^{\rm obs}$ that appears in eq.~(\ref{sigexp}) is obtained after applying 
detector- and analysis-dependent (lepton) cuts on a very complex final
state. The rescaling of $N^{\rm obs}$ by $1/A_W$ allows one to relate this
quantity to the relevant $W$ cross section; in other words, acceptance
corrections provide an estimate of the number of undetected events. The
acceptances need therefore be computed with a program that is able to reliably
describe in full details the final state emerging from $W$ production, which
typically means an event generator. The following issues then arise: how
accurately can these acceptance corrections be calculated? Does the accuracy
of the acceptance calculation match the intrinsic accuracy of the theoretical
prediction for the $W$ cross section? Generally speaking, the answer to the
latter question is negative: the very high accuracy of the theoretical
computations of the $W$ cross section is due to the fact that these
computations are fully inclusive, which is not the case for the codes used to
obtain the acceptances. However, if acceptance cuts only involve the leptons
coming from the $W$ decay, then one may use one of the fixed-order
computations~\cite{Altarelli:1979ub}--\cite{Anastasiou:2003ds}, letting the
$W$ decay isotropically in its rest frame to get the final-state leptons. This
procedure may lead to large errors: although the $W$ distributions are
correctly predicted by these computations, the lepton distributions are not,
since the spin correlations are neglected between the leptons and the
initial-state partons. NLO results are available~\cite{Aurenche:1980tp} that
include such spin correlations, but analogous NNLO results are beyond current
capabilities. It should be stressed that the computations that include EW
corrections~\cite{Baur:2001ze,Dittmaier:2001ay} do include lepton spin
correlations.

The aim of this paper is that of assessing the accuracy to which 
acceptances for $W$ signals can be estimated at the Tevatron and
the LHC. We shall limit ourselves to considering the QCD effects 
that may change the computation of the acceptances as obtained
with standard parton shower Monte Carlos.

The paper is organized as follows: in sect.~\ref{sec:prel} we
introduce our conventions and notations; sects.~\ref{sec:LOvsNLO}
and~\ref{sec:unc} present the results for acceptances and
distributions, and in sect.~\ref{sec:concl} we give our
conclusions.

\section{Preliminaries\label{sec:prel}}
We focus on $W$ production at the Tevatron \ppbar\ collider
($\sqrt{S}=1.96$~TeV) and at the LHC ($pp$, $\sqrt{S}=14$~TeV), and
assume in all cases leptonic decays of the $W$ (for the sake of
definiteness, we shall consider $W\to e\nu$; lepton mass 
effects will be neglected throughout this paper). In each
case, we shall discuss two possible sets of experimental cuts,
selected to reflect realistic detector capabilities, and to better
illustrate how different physics effects have different
impacts on the acceptances depending on the event definition. 

For the Tevatron, we define the following cuts:
\beqn
&&\bullet~~{\rm\bf Cut~1:}\quad\quad\quad\quad\quad\quad
\ptel>20~\gev \; ,  \quad \abs{\etael}<1 \; , \quad\quad\quad\;\;
\met>20~\gev\,;
\phantom{aaaaaaaaaaa}
\label{Tevcuto}
\\
&&\bullet~~{\rm\bf Cut~2:}\quad\quad\quad\quad\quad\quad
\ptel>20~\gev \; ,  \quad 1< \abs{\etael}<2.5 \; , \quad
\met>20~\gev\,.
\phantom{aaaaaaaaaaa}
\label{Tevcutt}
\eeqn
In both cases, we identify $\met$ with the transverse momentum of the
neutrino; $\ptel$ and $\etael$ are the transverse momentum and the rapidity 
of the electron. The different rapidity ranges for the two cases mimic
typical selection cuts used by the Tevatron experiments, and provide a
useful separation between regions of the $W$ rapidity spectrum which
have different sensitivities to some of the sources of uncertainty, 
such as the PDFs. At the LHC we define instead:
\beqn
&&\bullet~~{\rm\bf Cut~1:}\quad\quad\quad\quad\quad\quad
\ptel>20~\gev \; ,  \quad \vert \etael \vert <2.5 \; , \quad
\met>20~\gev\,;
\phantom{aaaaaaaaaaaaaa}
\label{LHCcuto}
\\
&&\bullet~~{\rm\bf Cut~2:}\quad\quad\quad\quad\quad\quad
\ptel>40~\gev \; ,  \quad \vert \etael \vert <2.5 \; , \quad
\met>20~\gev\,.
\phantom{aaaaaaaaaaaaaa}
\label{LHCcutt}
\eeqn
In this case the selection with higher $\ptel$ threshold is mostly intended 
to provide an example of a cut which is very sensitive to the accuracy of
the theoretical computation. In addition, large values of $\ptel$ will 
be used in the LHC triggers, to cope with the huge inclusive-electron 
signal and background rates. 

We now define our theoretical calculations in more detail:
\begin{itemize}
\item LO: parton-level LO QCD; 
\item LO+\herwig: parton-level LO QCD, evolved through the
 \herwig\ shower~\cite{Corcella:2000bw}. No matrix-element corrections 
 to the parton shower~\cite{Seymour:1995df}--\cite{Miu:1999ju} 
 have been included, to preserve the LO nature of this step;
\item NLO: parton-level NLO QCD; 
\item MC@NLO: parton-level NLO QCD, merged with the \herwig\ parton
  shower as discussed in refs.~\cite{Frixione:2002ik,Frixione:2003ei}.
  Version 2.31 of MC@NLO is used. 
\end{itemize}
The LO parton-level computations have been performed with 
\ALPGEN\ \cite{Mangano:2002ea}. The NLO matrix elements of
ref.~\cite{Aurenche:1980tp} have been implemented in a 
fully-differential code according to the formalism of
refs.~\cite{Frixione:1995ms,Frixione:1997np}; by turning off
the ${\cal O}(\as)$ corrections, complete agreement has been found
with the results obtained with \ALPGEN\ for all the $W$ 
and lepton observables considered. All of the cases above 
include the spin correlations between the decay leptons and the 
partons entering the hard matrix elements. For our comparisons, we shall 
also consider the case in which spin correlations are turned off, an option 
implemented by simply letting the $W$ boson decay with a pure phase-space
distribution. All $W$-width effects are included, and we generate
events for which the dilepton pair has a mass within the range
$\mw-30\gw < m_{e\nu} < \mw+30\gw$. The production
rate outside this range is below $10^{-3}$ of the total.

Our input parameters are defined by tree-level electroweak
gauge invariance~\cite{Mangano:2002ea}
 with fixed input values for \mw, \mz\ and $G_F$:
\be
\mw=80.419~\gev\,, \quad \gw=2.048~\gev\,, \quad \sin^2\theta_W=0.222\,.
\ee
As a default PDF set for all the calculations we use the NLO set 
MRST2001~\cite{Martin:2001es}, with \mbox{\asmz=0.119}. The 
default scale choice is $\mur=\muf=\mu_0\equiv\sqrt{\mwt+\ptw^2}$. 

\section{Results: shower effects at LO and NLO\label{sec:LOvsNLO}}
In this section we compute acceptances and total and differential
cross sections, comparing the results of the four theoretical
approaches defined in the previous section. We shall emphasize
in particular the role of NLO corrections, and that of the shower
acting on top of the LO and NLO parton-level matrix elements.
The effects of neglecting the spin correlations for the $W$ 
decay products will also be considered here.

\begin{center}
\begin{figure}[t]
\epsfig{file=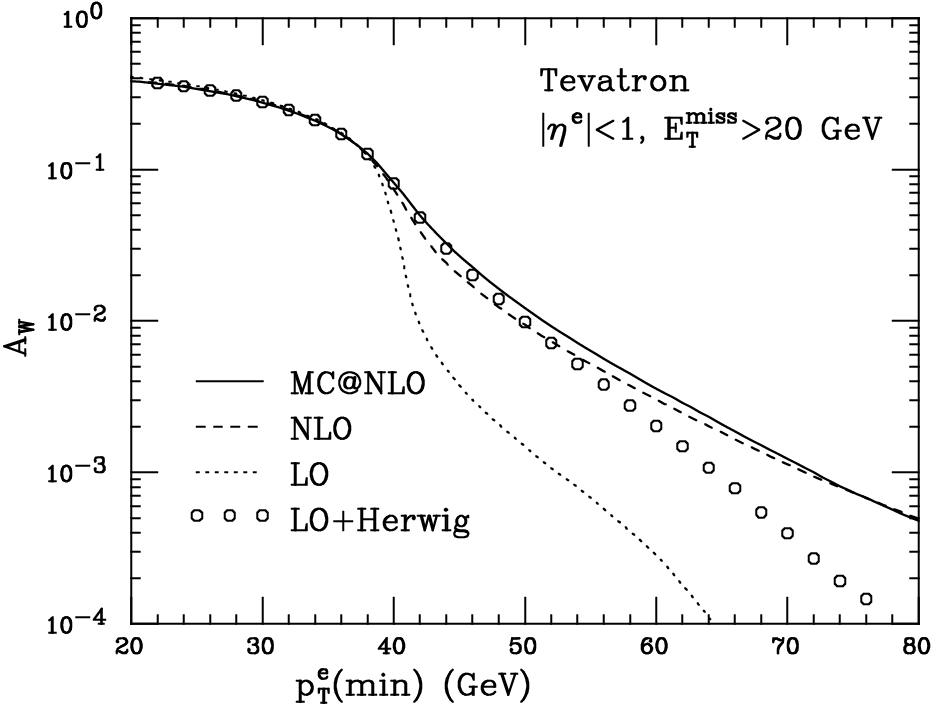,width=0.48\textwidth}\hfill
\epsfig{file=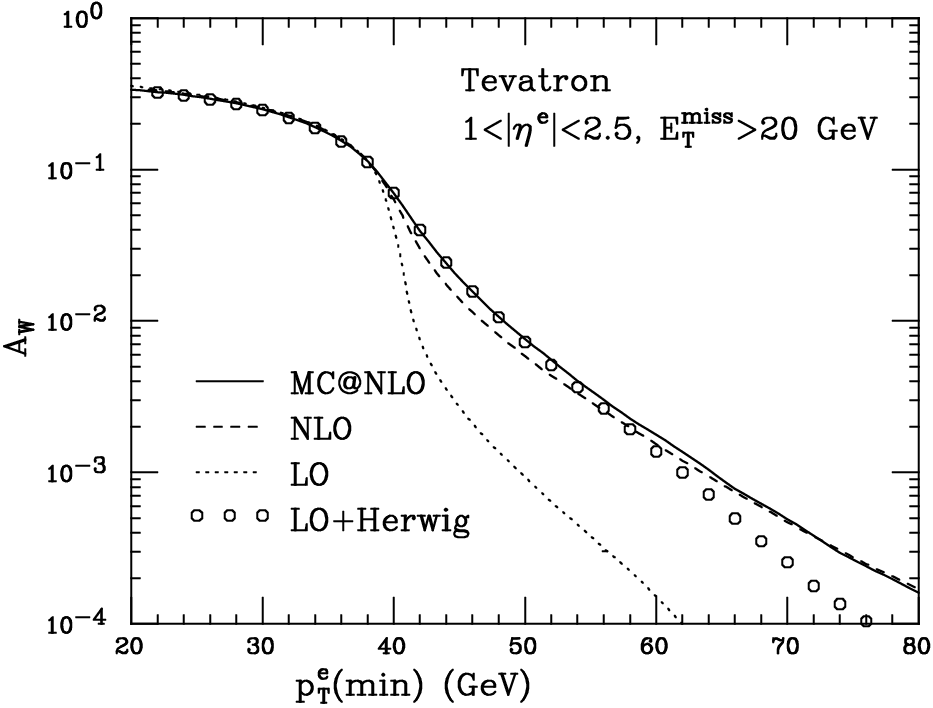,width=0.48\textwidth}
\ccaption{}{\label{fig:pttev} 
Acceptances as a function of the minimum electron transverse
momentum, for the two sets of cuts considered at the Tevatron.}
\end{figure}
\end{center}

\begin{center}
\begin{figure}[t]
\epsfig{file=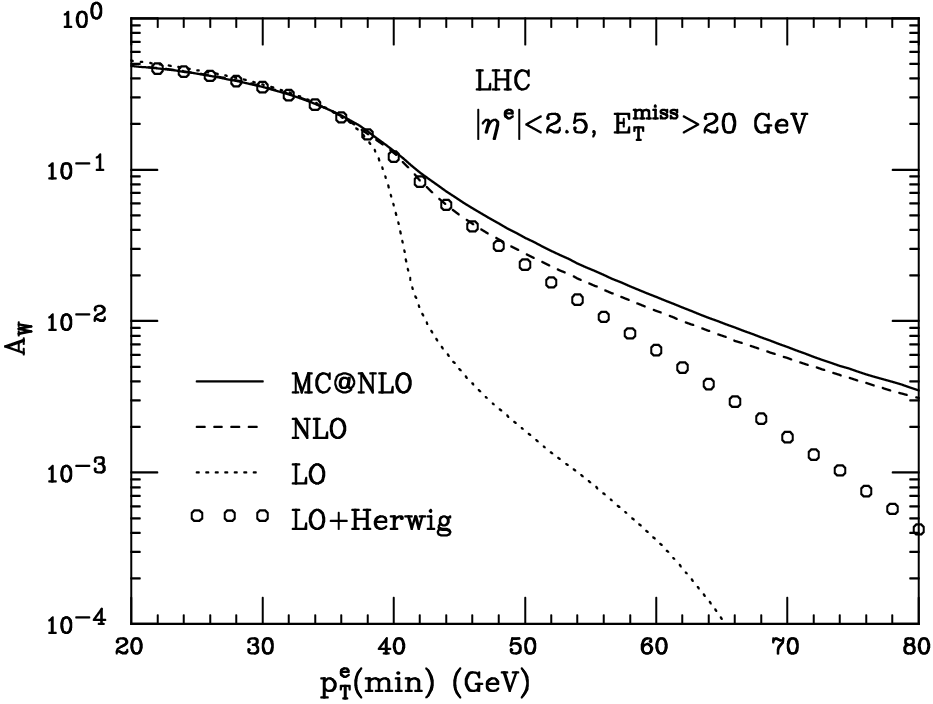,width=0.96\textwidth}
\ccaption{}{\label{fig:ptlhc}
Acceptances as a function of the minimum electron transverse
momentum, at the LHC.}
\end{figure}
\end{center}

We start with the lepton transverse momentum spectra, shown in
fig.~\ref{fig:pttev} and \ref{fig:ptlhc} for the Tevatron and the LHC,
respectively. The spectra are plotted in the form of acceptances as a
function of the minimum electron transverse momentum ($\ptel(\min)$), 
in events which satisfy the \etael\ and \met\ cuts:
\beq
A_W(\ptel(\min))=\frac{1}{\sigma^{(tot)}}\int_{\ptel(\min)}^{\sqrt{S}/2} 
d\ptel\,\frac{d\sigma}{d\ptel}({\rm cuts})\,,
\eeq
where $\sigma^{(tot)}$ is the total $W$ production cross section,
evaluated case by case in the appropriate scheme (LO or NLO).
Since at the LHC both cut~1 and cut~2 require the same constraints on \etael\
and \met, there is only one plot in this case. The four curves in each plot
correspond to the four theoretical computations introduced before.

There are clear differences among the four calculations in the
high-$\ptel$ region. The large difference between the LO and the other
results is due to the fact that the LO is the only case in which $\ptw=0$, 
which implies that $\ptel\le\mw/2$; thus, at the LO the high-$\ptel$ 
region is only populated by those events contributing to the 
tail of the $W$ mass spectrum. The addition of the parton shower improves
the situation, since the $W$ acquires a transverse momentum by recoiling
against the partons emitted by the shower. However, it is well known that
the $\ptw$ distribution which originates in this way is considerably
softer than that predicted at the NLO, since the shower lacks the hard 
\oas\ effects included in the NLO matrix elements. This fact is reflected
in the large differences at $\ptel \sim \mw$ between the LO+\herwig\ and the
NLO/MC@NLO predictions. Notice finally that while the NLO and 
MC@NLO results match quite well in the regions $\ptel\lsim \mw/2$
and $\ptel\gsim \mw$, the MC@NLO acceptance is larger than the
NLO one when $\mw/2 \lsim \ptel \lsim \mw$.
While the high-\pt\ region is not relevant to the determination of
the total cross section (since trigger thresholds are typically below
the $\mw/2$ value), it may play a role in the determination of the $W$
width, which is extracted from the shape of the high-transverse-mass spectrum
of the $\ell\nu$ pair~\cite{Affolder:2000mt,Abazov:2002xj}.

\begin{center}
\begin{figure}[t]
\epsfig{file=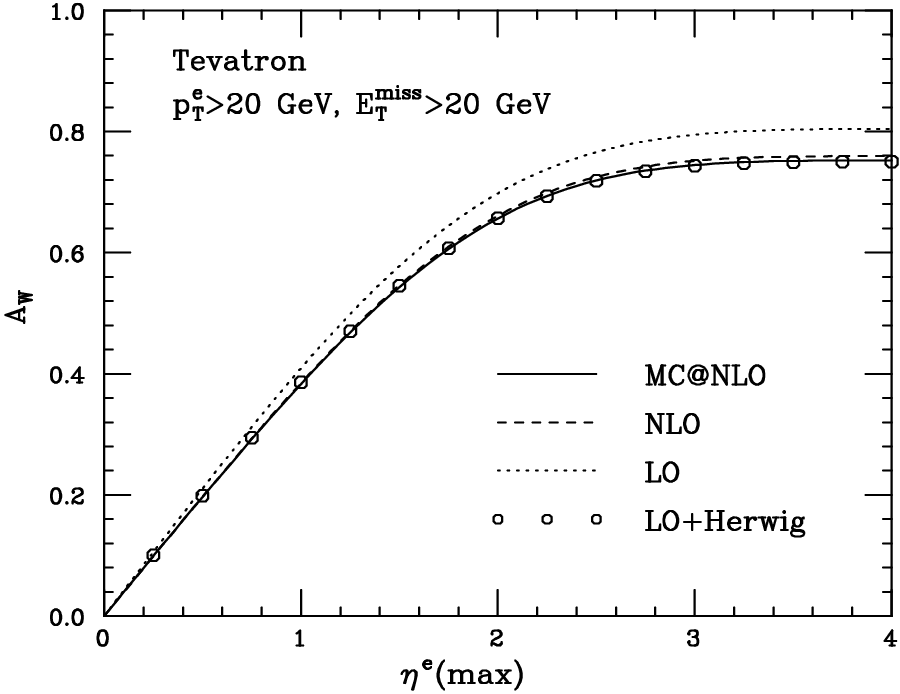,width=0.48\textwidth}\hfill
\epsfig{file=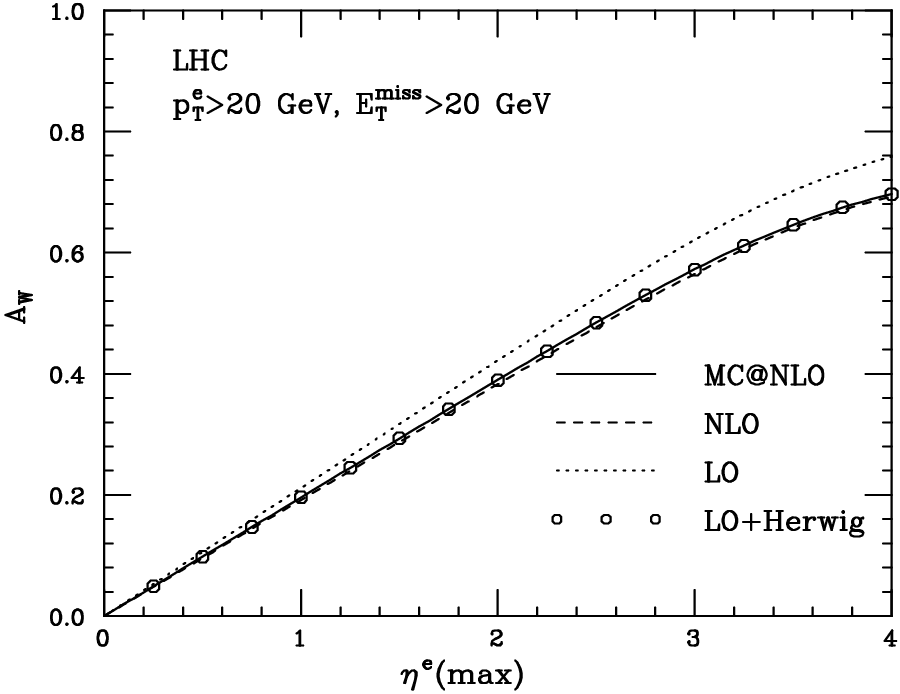,width=0.48\textwidth}
\ccaption{}{\label{fig:eta} 
Acceptances as a function of the maximum electron rapidity,
at the Tevatron (left panel) and the LHC (right panel), for the
smaller $\ptel$ cut.}
\end{figure}
\end{center}

The differences between the various approaches are much smaller in
the small-$\ptel$ region. In fig.~\ref{fig:eta} we plot the acceptance
defined as follows:
\beq
A_W(\etael(\max))=\frac{1}{\sigma^{(tot)}}\int_0^{\etael(\max)}
d\!\abs{\etael}\,\frac{d\sigma}{d\!\abs{\etael}}({\rm cuts})\,,
\eeq
as a function of the maximum electron rapidity ($\etael(\max)$), in events 
which satisfy the \mbox{$\ptel>20~\gev$} and the \met\ cuts; Tevatron (left 
panel) and LHC (right panel) results are presented. As can be inferred from 
fig.~\ref{fig:eta}, the relative behaviour of the various results at 
small $\ptel$'s shown in figs.~\ref{fig:pttev} and~\ref{fig:ptlhc} would 
not change had we integrated over different ranges in the electron rapidity. 
This implies that for measurements dominated by small $\ptel$'s the 
uncertainties on the acceptance corrections are basically independent 
of the electron rapidity range chosen.

In order to allow a closer comparison between the various theoretical
approaches, we present the results\footnote{The relative errors on our
results for total rates are $6\cdot 10^{-4}$ or smaller, and beyond the last 
digit reported for acceptances.} for the rates in tables~\ref{tab:tevtot}
and~\ref{tab:lhctot}, and for the acceptances in tables~\ref{tab:tevacc}
and~\ref{tab:lhcacc}, for the Tevatron and the LHC.
In the case of the two sets of cuts considered at the Tevatron, 
eqs.~(\ref{Tevcuto}) and~(\ref{Tevcutt}), the LO+\herwig, NLO, and MC@NLO
predictions are very close to each other, whereas the LO results differ by 
more than 5\%. It is particularly remarkable that the addition of the shower 
has a sizable effect on the LO result, while the NLO result remains 
essentially the same. From the left panel of fig.~\ref{fig:eta}, we 
see that the LO+\herwig\ and the MC@NLO predictions would be even closer to
each other, had we considered a larger $\etael$ range than that of cut 1.
It is worth noting that, by adding the shower to the LO matrix elements, 
there are two effects; the first one has already been mentioned, and consists 
in giving a non-zero $\pt$ to the $W$. The second effect is due to the
fact that, by the backward showering of the partons which enter the
LO matrix elements, one may end up with one or two gluons emerging
from the colliding hadrons (whereas at the LO only the $q\bar{q}$ 
combination is possible). This fits nicely into the picture of perturbative
QCD corrections; in fact, at the NLO both the $q\bar{q}$ and the 
$qg+\bar{q}g$ partonic initial states contribute to the results.
At the Tevatron the former effect is by far the dominant one, as 
can be verified by computing the acceptances for cut 1 and 2 at the
NLO and considering only the $q\bar{q}$ contributions\footnote{Such 
contributions are scale and scheme dependent; however, this can be neglected
for the sake of the present qualitative argument.}, which turn out to be very
close to the full NLO results. 
On the other hand, in the case of cut 1 at the LHC, the
$q\bar{q}$ contribution to the NLO acceptance is about 4\% larger than
the full NLO result. Thus, both effects play a role in the nice agreement 
between the LO+\herwig\ and the MC@NLO results for cut 1 at the LHC, shown in 
table~\ref{tab:lhcacc}. As already observed in figs.~\ref{fig:pttev} 
and~\ref{fig:ptlhc}, this situation changes when the $\ptel$ threshold 
is increased -- see the results relevant to cut 2 in 
table~\ref{tab:lhcacc}: the MC@NLO prediction is 9\% (3\%) larger
than that of LO+\herwig\ (NLO). Here, the difference between MC@NLO and
LO+\herwig\ is essentially due to the lack of hard emissions in the
latter -- 40~GeV is large enough for the collinear approximation built
into the shower to start failing. The difference between MC@NLO and NLO
has a different origin: in the parton-level LO computation at a fixed
$\mw$, $\ptel$ cannot assume values larger than $\mw/2$; this 
implies the possible presence of large logarithmic terms, that arise
to all orders beyond the leading one in perturbation theory, and that 
can be effectively resummed by the shower in MC@NLO. The impact of these 
logs is less important as one moves away from the threshold, as can be 
seen from figs.~\ref{fig:pttev} and~\ref{fig:ptlhc}.

The inclusion of NLO matrix elements into a parton shower framework
renders the computation of the acceptances by MC@NLO intrinsically
more reliable than that performed with a standard Monte Carlo event
generator. However, one may wonder whether the accuracy thus obtained
is sufficient in the context of an NNLO analysis. From the discussion
given above, it seems indeed so. In fact, no qualitatively new
kinematic effects appear at the NNLO with respect to the NLO; as
far as the computation of the $W$ acceptance is concerned, it is
irrelevant whether the $W$ recoils against one or two hard partons
(the same would not be true were we interested in the $W$+jet system).
The $\ptel=\mw/2$ boundary is treated by MC@NLO to all orders, thus
including NNLO effects. The partonic initial states that appear for
the first time at the NNLO, such as $gg$, have a very modest impact
on $W$ distributions~\cite{Anastasiou:2003ds}, and are anyhow included 
in MC@NLO through backward showering. Thus, we expect the acceptances 
computed with MC@NLO to be fairly similar to those that could be
computed if we knew how to merge NNLO matrix elements with parton showers.
{\renewcommand{\arraystretch}{1.2}
\begin{table}
\begin{center}
\begin{tabular}{l|cccc} \hline
               & LO  & LO+HW & NLO   & MC@NLO \\
TeV Total      & 2220     & 2220  & 2679  & 2679   \\
TeV cut 1      &  907     & 856   & 1031  & 1027   \\
TeV cut 2      &  790     & 738   & 911   & 900  \\
\hline 
\end{tabular}                                                                 
\ccaption{}{\label{tab:tevtot} Predictions for the $W$ production
  rates (in pb) at the Tevatron, with and without the selection cuts 
  defined in eqs.~(\ref{Tevcuto}) and~(\ref{Tevcutt}).}
\end{center}                                         
\end{table} }
{\renewcommand{\arraystretch}{1.2}
\begin{table}
\begin{center}
\begin{tabular}{l|cccc} \hline
               & LO       & LO+HW & NLO    & MC@NLO \\
LHC Total      & 18270    & 18300 & 20900  & 20900  \\
LHC cut 1      & 9580     & 8861  & 9970   & 10125  \\
LHC cut 2      & 1060     & 2230  & 2699   & 2776 \\
\hline 
\end{tabular}                                                                 
\ccaption{}{\label{tab:lhctot} Predictions for the $W$ production
  rates (in pb) at the LHC, with and without the selection cuts 
  defined in eqs.~(\ref{LHCcuto}) and~(\ref{LHCcutt}).}
\end{center}                                         
\end{table} }
{\renewcommand{\arraystretch}{1.2}
\begin{table}
\begin{center}
\begin{tabular}{l|cccc} \hline
               & LO & LO+HW & NLO    & MC@NLO \\
TeV cut 1      &  0.409  & 0.386 & 0.385 & 0.383 \\
TeV cut 2      &  0.356  & 0.333 & 0.340 & 0.336 \\
\hline 			  
\end{tabular}                                                                 
\ccaption{}{\label{tab:tevacc} Acceptances for the various
  cuts, at the Tevatron.}
\end{center}                                         
\end{table} }
{\renewcommand{\arraystretch}{1.2}
\begin{table}
\begin{center}
\begin{tabular}{l|cccc} \hline
               & LO      & LO+HW  & NLO   & MC@NLO \\
LHC cut 1      &  0.524  & 0.484  & 0.477 & 0.485 \\
LHC cut 2      &  0.058  & 0.122  & 0.129 & 0.133 \\
\hline 			  
\end{tabular}                                                                 
\ccaption{}{\label{tab:lhcacc} Acceptances for the various
  cuts, at the LHC.}
\end{center}                                         
\end{table} }

Given the fact that, as shown before for $\ptel>20~\gev$, MC@NLO and NLO
give similar results, one may take a different attitude, and use 
parton-level NNLO results to compute the acceptances. If the cuts
chosen do not select {\em only} the region of small $\ptw$ (which 
is not reliably predicted by any fixed-order computation), the result
of ref.~\cite{Anastasiou:2003ds} gives access to the full $W$ kinematics.
Unfortunately, no NNLO computation includes lepton spin correlations.
These correlations are irrelevant if one is interested in the distributions
of the $W$, but are important if one needs to apply cuts on lepton
variables. To document this, we present in table~\ref{tab:accnospin}  
the acceptance results obtained by switching the spin correlations off 
(namely assuming flat, phase-space decays of the $W$); we compare the LO, 
NLO, and MC@NLO results with the analogous ones obtained with full spin 
correlations, already reported before. Apart from the case of cut 1 at
the Tevatron, the effects are very large, with shifts of up to 15\%. 
NLO and MC@NLO are in general close to each other, but no clear pattern
emerges when going from LO to NLO; as it should be expected, the LO to NLO
ratio depends on the electron rapidity range considered. We thus conclude
that, lacking the full information on lepton spin correlations, present 
NNLO results can only give rough estimates of the acceptances.
{\renewcommand{\arraystretch}{1.2}
\begin{table}
\begin{center}
\begin{tabular}{l|ccc|ccc} \hline
& \multicolumn{3}{c|}{Tevatron} & \multicolumn{3}{c}{LHC} \\ 
                & LO    & NLO    & MC@NLO & LO    & NLO   & MC@NLO \\
\hline
Cut 1           & 0.409 & 0.385 & 0.383 & 0.524 & 0.477 & 0.485\\
Cut 1, no spin  & 0.413 & 0.394 & 0.394 & 0.553 & 0.510 & 0.515\\
\hline
Cut 2           & 0.356 & 0.340 & 0.336 & 0.058 & 0.129 & 0.133\\
Cut 2, no spin  & 0.389 & 0.374 & 0.370 & 0.075 & 0.150 & 0.157\\
\hline 			  
\end{tabular}                                                                 
\ccaption{}{\label{tab:accnospin}  
Effect of spin correlations on acceptances for the various
cuts, at the Tevatron and the LHC.}
\end{center}                                         
\end{table} }

\begin{center}
\begin{figure}[t]
\epsfig{file=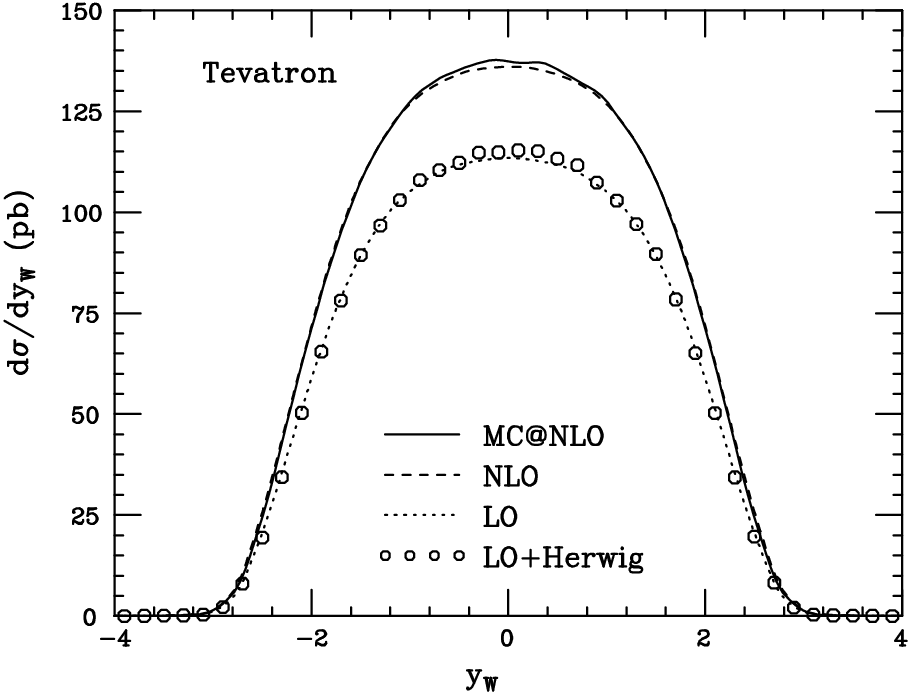,width=0.48\textwidth}\hfill
\epsfig{file=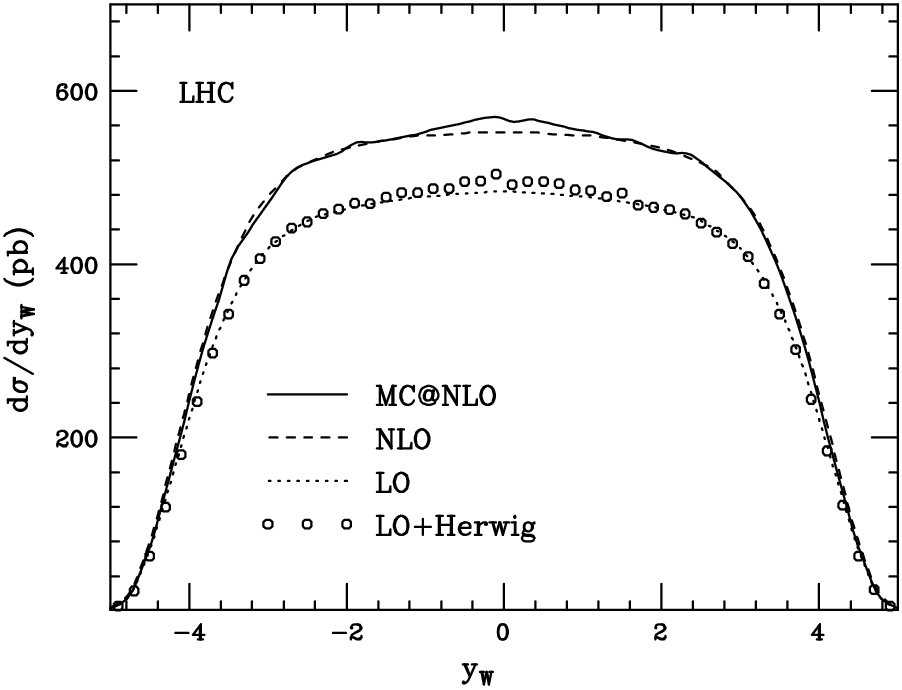,width=0.48\textwidth}
\ccaption{}{\label{fig:Wy} 
Fully-inclusive $W$ rapidity distribution, at the Tevatron (left panel)
and the LHC (right panel).}
\end{figure}
\end{center}

\begin{center}
\begin{figure}[t]
\epsfig{file=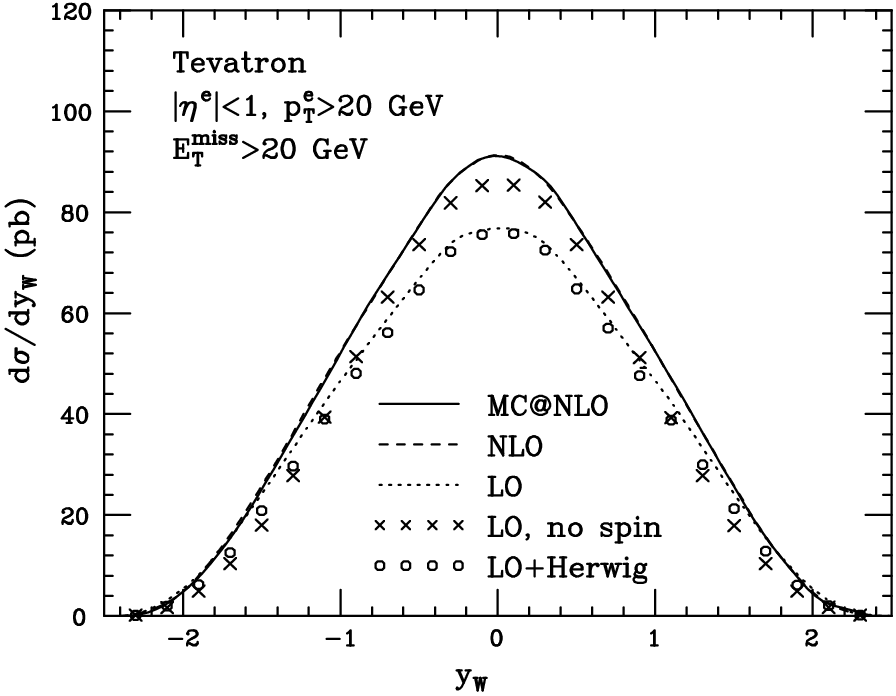,width=0.48\textwidth}\hfill
\epsfig{file=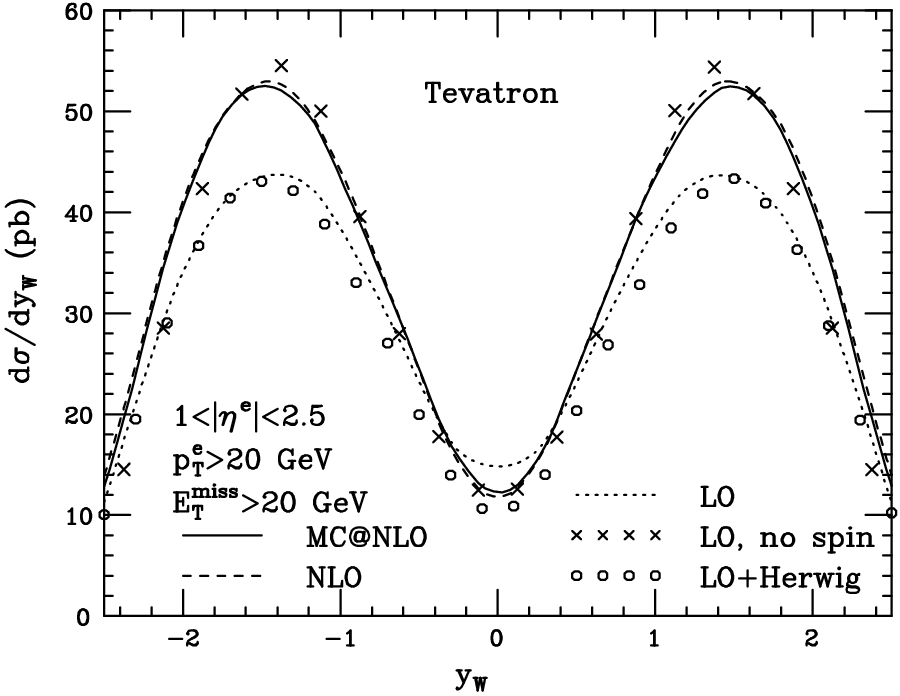,width=0.48\textwidth}
\ccaption{}{\label{fig:Wycut_tev} 
$W$ rapidity distribution at the Tevatron, with lepton cuts.}
\end{figure}
\end{center}

\begin{center}
\begin{figure}[t]
\epsfig{file=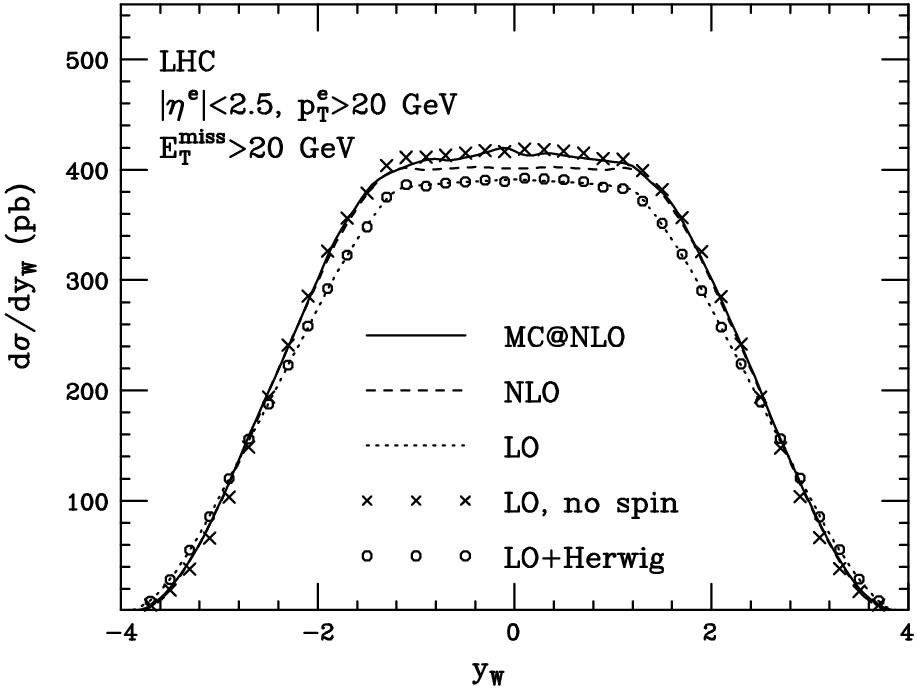,width=0.96\textwidth}
\ccaption{}{\label{fig:Wycut_lhc} 
$W$ rapidity distribution at the LHC, with lepton cuts.}
\end{figure}
\end{center}

We can also consider quantities that are less inclusive than
acceptances, such as the rapidity of the $W$ boson ($\yw$). It has been shown
in ref.~\cite{Anastasiou:2003ds} that the \yw\ spectrum at NNLO can be 
very accurately reproduced by rescaling the NLO distribution with the
appropriate K factor; interestingly, the rescaled LO distribution
is {\em not} a good approximation of the full NLO distribution.
The arguments given before on MC@NLO imply that, by rescaling with
the K factor the $\yw$ distribution predicted by MC@NLO, we should
get a good approximation of the true NNLO+shower prediction.
In fig.~\ref{fig:Wy} we show the fully inclusive \yw\ spectra for the
Tevatron and the LHC. We notice that the inclusion of the shower into
both the LO and NLO calculations leads to a slightly more central
production, in particular at the LHC. After imposing lepton selection
cuts, this effect is reduced at the Tevatron (fig.~\ref{fig:Wycut_tev}), 
but remains clearly visible at the LHC at the NLO (fig.~\ref{fig:Wycut_lhc}). 
In analogy to what done in table~\ref{tab:accnospin}, we also include the 
predictions obtained at the LO by switching the spin correlations off (the
curves are those labelled ``LO, no spin''), which result in significant
changes in the shapes of the distributions.


\section{Results: PDF and scale uncertainties\label{sec:unc}}
In this section we study the sensitivity of the acceptances
to the uncertainties affecting the PDF sets and to the choices
of renormalization/factorization scales. 

To assess the PDF uncertainty we have used the 30 MRST2001E 
sets~\cite{Martin:2002aw}, using the prescription for asymmetric
errors proposed in ref.~\cite{Sullivan:2002jt} (modified tolerance
method). As discussed before, our best estimates of the acceptances are
obtained with MC@NLO.  Given the results relevant to cuts 1 and 2, we expect
the uncertainties relative to the central value to be very similar when
computed with MC@NLO or with NLO. Thus, we shall restrict ourselves here to
the parton-level NLO computations, which are somewhat faster to perform 
than MC@NLO's.
Using the default MRST2001E tolerance value of $T=\sqrt{50}$ (see below for
further discussions on this point), we obtain the following results for 
the Tevatron:
\ba
\sigma({\rm NLO}) &=& 2679 {+26 \atop -43}\;\; {\rm pb} 
\label{uncone}
\\
A_W(\mbox{cut 1}) &=& 0.3848 {+0.0020 \atop -0.0039} 
\\
A_W(\mbox{cut 2}) &=& 0.3402 {+0.0028 \atop -0.0013}
\ea
and for the LHC:
\ba
\sigma({\rm NLO}) &=& 20900 {+318 \atop -474}\;\; {\rm pb}
\\
A_W(\mbox{cut 1}) &=& 0.4770 {+0.0048 \atop -0.0049} 
\\
A_W(\mbox{cut 2}) &=& 0.1292 {+0.0007 \atop -0.0027}
\label{uncsix}
\ea
To compute the uncertainties shown in these equations, we considered
the pulls with respect to the results obtained with the $n=0$ MRST2001E
set. Although the $n=0$ set is very similar to the default set of MRST2001, 
the results obtained with the two are not identical for the cross sections
(with the $n=0$ set, we get 2673~pb and 20815~pb at the Tevatron and the 
LHC respectively) -- they are identical for the acceptances; however, 
these differences being less than 0.5\%, we associated the uncertainties 
computed with the $n=0$ set with the cross section results relevant to 
the default set of MRST2001.

{\renewcommand{\arraystretch}{1.2}
\begin{table}
\begin{center}
\begin{tabular}{l|cc|cc|cc} \hline
   & \multicolumn{2}{c|}{$\mu=\mu_0/2$} &
 \multicolumn{2}{c|}{$\mu=\mu_0$} &
\multicolumn{2}{c}{$\mu=2\mu_0$} \\
            & NLO    & MC@NLO  & NLO    & MC@NLO  & NLO    & MC@NLO \\
\hline
TeV cut 1   & 0.382 & 0.384     & 0.385 & 0.383 & 0.388 & 0.385 \\
TeV cut 2   & 0.339 & 0.335     & 0.340 & 0.336 & 0.342 & 0.335 \\
\hline 			  
\end{tabular}                                                                 
\ccaption{}{\label{tab:mutevacc}  
Scale dependence of the acceptances, Tevatron.} 
\end{center}                                         
\end{table} }
We remind the reader that the Hessian method~\cite{Pumplin:2001ct}, as
it has been originally proposed, returns symmetric uncertainties for a
given observable. These uncertainties are therefore independent
of the central value for the observable considered, at variance
with what obtained in eqs.~(\ref{uncone})--(\ref{uncsix}) with the
prescription of ref.~\cite{Sullivan:2002jt}. Following
ref.~\cite{Pumplin:2001ct} we would have obtained 
\beq
\Delta\left(\sigma({\rm NLO})\right)=32\;{\rm pb},\;\;\;\;
\Delta\left(A_W(\mbox{cut 1})\right)=0.0027,\;\;\;\;
\Delta\left(A_W(\mbox{cut 2})\right)=0.0018,\;\;\;\;
\label{usymmTev}
\eeq
at the Tevatron, and
\beq
\Delta\left(\sigma({\rm NLO})\right)=386\;{\rm pb},\;\;\;\;
\Delta\left(A_W(\mbox{cut 1})\right)=0.0047,\;\;\;\;
\Delta\left(A_W(\mbox{cut 2})\right)=0.0015,\;\;\;\;
\label{usymmLHC}
\eeq
at the LHC. It is reassuring that these results are in overall agreement
with those shown in eqs.~(\ref{uncone})--(\ref{uncsix}).
We notice that both at the Tevatron and the LHC the relative uncertainty on
the acceptance is approximately half the size of the uncertainty on the total
rate, and at the per cent level. Since the impact of the PDFs on the
acceptance is mostly due to the \etael\ cuts, which reflect the \yw\
distributions, we expect that accurate measurements of the \etael\ spectra
will allow to reduce this uncertainty even further once the data will be
available.

\begin{table}
\begin{center}
\begin{tabular}{l|cc|cc|cc} \hline
   & \multicolumn{2}{c|}{$\mu=\mu_0/2$} &
 \multicolumn{2}{c|}{$\mu=\mu_0$} &
\multicolumn{2}{c}{$\mu=2\mu_0$} \\
           & NLO    & MC@NLO  & NLO    & MC@NLO  & NLO    & MC@NLO \\
\hline
LHC cut 1  & 0.475 & 0.485  & 0.477 & 0.485 & 0.478 & 0.484 \\
LHC cut 2  & 0.130 & 0.134  & 0.129 & 0.133 & 0.125 & 0.132 \\
\hline 			  
\end{tabular}                                                                 
\ccaption{}{\label{tab:mulhcacc}  
Scale dependence of the acceptances, LHC.}
\end{center}                                         
\end{table} 
We finally point out that the theoretical picture underlying the
treatment of PDF uncertainties is far from being established. Although
within the Hessian method one formally arrives at the definition of the
$1\sigma$ error band, in practice the combined effect of the failure of
some of the theoretical approximations involved, and of difficulties in 
the treatment of the correlations between experimental errors, implies 
the necessity of dropping the rigorous $1\sigma$ considerations. At this
point, one is forced to introduce an {\em arbitrariness} in the procedure,
parametrized in terms of a single parameter (the tolerance) $T$, which is the
maximum allowed of the $\Delta\chi^2$ variation w.r.t. the parameters of the
best PDF fit. The MRST2001E~\cite{Martin:2002aw} and
CTEQ6~\cite{Pumplin:2002vw} sets assume $T=\sqrt{50}$ and $T=10$
respectively. On this basis alone, and barring the other differences between
the parametrizations of refs.~\cite{Martin:2002aw} and~\cite{Pumplin:2002vw},
with the latter choice for $T$ the uncertainties of
eqs.~(\ref{uncone})--(\ref{usymmLHC}) would have been a factor of $\sqrt{2}$
larger. Furthermore, it has been argued that the Lagrange multiplier
method~\cite{Stump:2001gu} may be better suited if one is interested in
specific observables, such as the ones considered in this paper. In
ref.~\cite{Martin:2002aw} the PDF uncertainty affecting the $W$ cross section
at the LHC, computed according to the Lagrange multiplier method, has been
found to be marginally larger than that computed with the Hessian method. We
conclude that, although the results of eqs.~(\ref{uncone})--(\ref{uncsix}) are
based on some assumptions that will need further theoretical considerations,
they can be considered as reliable estimates, perhaps up to a factor of 1.5,
of the PDF uncertainties.

In tables~\ref{tab:mutevacc} and~\ref{tab:mulhcacc} we finally present 
the results for the scale dependence of the acceptances at the Tevatron 
and the LHC. We identify the factorization and renormalization scales, 
and set them equal to $r_\mu \mu_0$, with $r_\mu=1/2,1$ and 2. 
The uncertainty at NLO is of the order of 1--2\%, depending on the
cuts (the largest variation being obtained for cut~2 at the LHC). It
is reduced to below 1\% with MC@NLO. Although an independent variation
of $\mur$ and $\muf$ would lead to larger uncertainties, these results 
suggest a good stability w.r.t. to the addition of NNLO corrections
(as shown explicitly in ref.~\cite{Anastasiou:2003ds} for the case
of fully inclusive $\yw$ distributions), and point towards an 
improved scale dependence of the full NLO+shower result. This
behaviour is typical of most of the matched computations, which
combine the matrix elements computed to a given order in perturbation
theory with the resummation of large logarithmic terms.

\section{Conclusions\label{sec:concl}}
We summarize here the main conclusions of our study.
\begin{itemize}
\item In the case of lepton \pt\ thresholds at 20~GeV, 
 the addition of the shower corrections to the parton-level LO
 calculation has a large effect on the acceptances. On the contrary,
 the addition of the shower changes the NLO parton-level result by
 only 1\%, both at the Tevatron and at the LHC and over the full
 rapidity range  $\vert \etael \vert < 2.5$. At the LO, the effect 
 of shower corrections increases with the lepton \pt\ threshold; at
 the NLO, it first increases (it is about 3\% for the 40~GeV threshold 
 at the LHC, and it is larger than 10\% around 50~GeV), and then
 decreases again when the threshold moves towards $\mw$.
\item A major role in the overall acceptance is  played by spin
  correlations. Only their inclusion can guarantee a solid estimate
  of the acceptance. No clear pattern of evolution of the spin
  correlations emerges when going from LO to NLO, indicating that no
  obvious guess can be made on the impact of spin correlations at the
  NNLO level. As a result, only when spin correlations will be
  included in the NNLO calculation it will be possible to use this
  improved result for solid acceptance predictions at the parton level.
\item The scale dependence of the acceptance is at a level of 1\% or
  less, suggesting that the NLO approximation is stable relative to
  the addition of NNLO corrections. This is consistent with the
  observation of ref.~\cite{Anastasiou:2003ds} that the shape of the 
  fully inclusive rapidity distribution of the $W$ boson is not altered 
  by NNLO effects. Since the \yw\ distribution is one of the main elements
  determining the rapidity acceptance for the final-state charged lepton, it
  is therefore reasonable to assume that this conclusion survives the 
  presence of analysis cuts.
\item Current PDF uncertainties affect the calculation of the
  acceptance at the level of 1\%.
\end{itemize}
We conclude that the tools currently available (parton-level
NLO plus the parton shower, \`{a} la MC@NLO) should be sufficient to
guarantee an overall theoretical uncertainty on $W$ acceptances 
due to QCD effects at the level of 2\%, with possible improvements coming from
an in-situ monitoring of the rapidity distributions, which should reduce 
the PDF uncertainties. In addition to the QCD effects, the known EW 
corrections both to the short-distance matrix elements and to the definition
of the lepton energy and isolation will need to be included in any solid
experimental estimate of the total $W$ cross section. This overall accuracy
well matches the best theoretical estimates of the total $W$ cross
section, based on NNLO QCD and NLO EW calculations. This opens the way for
tests of QCD in hadronic collisions at the per cent level, and for
high-precision luminosity monitors based on large-rate and high-\pt\
observables.

\end{document}